# Electride States and Superconductivity in Dense Potassium Carbides

Jiance Sun[1], Ting Zhong[1], and Shoutao Zhang[1,*]

[1]State Key Laboratory of Integrated Optoelectronics and Key Laboratory of UV-Emitting Materials and Technology of Ministry of Education, School of Physics, Northeast Normal University, Changchun 130024, China

**Abstract**

Metal carbides have attracted a great deal of attention due to their diverse geometric motifs, remarkable physicochemical properties, and widespread practical applications. However, there is still a lack of systematic understanding regarding the phase diagram of pressurized potassium carbide. Employing first-principles swarm-intelligence structure prediction approach, we comprehensively explore the binary potassium-carbon phases under compression and identify a series of new K-rich and C-rich stoichiometric compounds. Among them, $K_7C$ with high K concentration manifests a monoclinic structure with space group *C*2/*m* and is predicted to be an electride with zero-dimensional (0D) interstitial electrons. C-abundant KC has an orthorhombic configuration with symmetry *Imma*, with the carbon atoms arranged in a zigzag pattern. Strikingly, $KC_3$, belonging to a monoclinic *C*2/*m* structure, possesses the highest carbon content and features a crumpled honeycomb carbon layer. Furthermore, calculations of electron-phonon coupling reveals that $K_7C$ is a 0D electride superconductor with a transition temperature ($T_c$) of 0.6 K at a pressure of 25 GPa. By contrast, KC exhibits a maximum $T_c$ of 21.4 K at 25 GPa, which is primarily attributed to the robust coupling between low-frequency K- and C-derived phonon modes and C 2*p* electrons at the Fermi level. In addition, $KC_3$ is calculated to have a $T_c$ value of 6.7 K at 25 GPa. This study provides valuable insights into K-C compounds and broadens the diversity of metal carbide superconductors.

## 1. Introduction

Superconductivity has always been one of the most fascinating research areas in condensed-matter physics, especially in the search for high-temperature and even room-temperature superconductors and the discovery of their origins[1–3]. According to Bardeen-Cooper-Schrieffer (BCS) theory, light-element compounds always have a high Debye temperature, which favors high-temperature superconductivity[4,5]. Recently, therefore, near-room-temperature superconductivity has been found in pressurized metal hydrides by following this theory[6,7]. However, instability has become the most significant hindrance to the development of hydride superconductors[8,9]. Consequently, in order to explore the high-temperature superconductors at low pressure, researchers changed their attention to other compounds made of light elements, such as nitrides, borides, carbides, boron carbides and so on[10–12].

[*]E-mail: zhangst966@nenu.edu.cn

Carbon is the sixth element in the periodic table and widely exits in our life. Due to diverse hybridization forms of *sp*, $sp^2$ and $sp^3$ between carbon atoms, the most abundant structures and morphologies can be formed such as diamond, graphite, nanotube and fullerene[13,14]. Notably, carbon-based materials have been explored experimentally and theoretically and proved to be potential high-temperature superconductors, playing significant roles in energy, information, medical, and other fields. For zero-dimensional (0D) carbon structures, the unconventional superconductivity of alkali metal–doped fullerene crystals has been extensively studied. For instance, Hebard *et al.* discovered the superconductivity of 18 K in the K-doped $C_{60}$[15]. Subsequently, researchers found that as the size of alkali metal cations increased, the superconductivity was enhanced. In Cs-doped $C_{60}$, the $T_c$ was predicted to reach 38 K[16]. Some clathrates with similar structures to $C_{60}$ have also been studied. For instance, it was found that the $T_c$ of Na-doped $C_{20}$ near 55 K[17], and researchers predicted that as the size of carbon cage increase, the superconductivity decreased[18].

For one-dimensional (1D) carbon structures, the superconductivity of carbon nanotubes has aroused great interest among a wide range of researchers. For instance, single-walled carbon nanotubes (SWCNTs) with 4 Å length exhibit 15 K superconductivity[19] .Subsequently, B-doped SWCNTs show 12 K superconductivity[20]. For two-dimensional (2D) system, graphite intercalation compounds (GICs) are crucial research objects. Their characteristic is that some atoms or molecules are inserted into interlayer spaces of graphene. In 1965, Hannay *et al.* observed the superconductivity in graphite intercalated with alkali metal K, Rb, and Cs, which is the earliest discovered carbon-based superconductor, but the $T_c$ is lower than 1 K[21]. Subsequently, Weller *et al.* observed superconductivity at 11.5 K and 6.5 K in GICs $CaC_6$ and $YbC_6$, respectively[22]. It is worth noting that $CaC_6$ has the highest $T_c$ among all the metal intercalated graphite compounds. In Li-modified monolayer graphene, angle-resolved photoemission spectroscopy technique was employed to reveal that the $T_c$ was only about 5.9 K[23]. Recently, unconventional superconductivity below 2 K has been observed in magic-angle graphene, which has aroused great interest in carbon-based materials[24,25]. Three-dimensional (3D) carbon structures also show outstanding superconductivity. It is well known that since all the valence electrons of diamond form covalent bonds, there are no free electrons available for conduction, and thus diamond is a typical insulator. However, the researchers found experimentally that B-doped diamond exhibit 4 K superconductivity[26], and the $T_c$ increased to 55 K as the B concentration rose[27]. Ding *et al.* found that H-doped diamond exhibit superconductivity exceeding 100 K at 5 GPa, which is the highest record among superhard materials[28]. The exotic superconductivity has also been observed in 3D carbon structures formed by coupling of carbon cages. For instance, the superconductivity of $T_c$ = 55 K was observed in a B-doped Q-carbon structure[29]. In the clathrate carbon structures of F-doped $B_{34}$ cages, the $T_c$ was theoretically predicted near 77 K[30]. Furthermore, it is predicted that the $T_c$ of metastable $NaC_6$, $MgC_6$, and $CsC_6$ with a sodalite-like structure composed of coupled carbon cages can reach over 100 K[31,32]. It can be seen that binary metal carbides are promising candidates with favorable superconducting properties. However, there is still a lack of research on high-pressure phase diagrams, chemical bonding characteristics, and superconducting behaviors under non-stoichiometric ratios of binary potassium-carbon carbides. Therefore, it is imperative to explore pressure-induced novel K-C compounds with unconventional stoichiometries, diverse motifs, unexpected electride states, and superior superconductivity.

In this work, systematic theoretical calculations and structure predictions were performed on the potassium-carbon system at pressures of 1 atm, 10, 25, 50, 100, 200, and 300 GPa. A variety of previously unreported metallic K-C crystal structures were identified, namely $K_8C$, $K_7C$, $K_4C$, $K_3C$, $K_2C$, KC, $KC_2$ and $KC_3$. Notably, these crystal structures feature diverse carbon atomic configurations, including isolated carbon atoms, carbon dimers, zigzag carbon chains, carbon planes and folded carbon layers. Furthermore, the newly discovered K-C compounds exhibit excellent metallic properties and can serve as potential superconducting candidates. More importantly, the *C*2/*m* phase of $K_7C$ possesses a distinctive physical characteristic where the electride state and superconducting state coexist. This study not only expands the family of metal carbide materials, but also deepens the understanding of their structures and physical properties.

## 2. Computational methods

To screen reasonable and stable structures of K-C compounds, this study systematically explores and searches crystal structures by adopting the CALYPSO structure prediction technique and first-principles calculation method[33,34]. Compared with conventional structure search methods, the CALYPSO approach exhibits outstanding advantages in discovering stable and metastable crystal structures under fixed chemical composition constraints, which provides efficient support for the exploration of novel structures[35-39].

On this basis, density functional theory (DFT) serves as the fundamental theoretical framework. The geometric optimization of crystal structures and total energy calculations considering spin polarization are implemented via the VASP code[40,41]. The Perdew-Burke-Ernzerhof (PBE) functional within the generalized gradient approximation (GGA) is adopted to describe exchange-correlation interactions and guarantee computational accuracy. The projector augmented wave (PAW) method is employed to characterize interactions between ions and electrons. The valence electron configurations are set as $3s^2 3p^6 4s^1$ for K pseudopotential and $2s^2 2p^2$ for C pseudopotential[42]. An energy cutoff of 800 eV is applied for wavefunction expansion to ensure energy convergence. The Monkhorst-Pack scheme is utilized for *k*-point sampling in the first Brillouin zone, with a *k*-point spacing of $2\pi\times0.025$ Å$^{-1}$. Phonon properties are calculated using the Phonopy package based on the finite displacement and supercell methods to accurately obtain phonon vibration modes and relevant parameters[43,44].

Electron-phonon coupling (EPC) calculations are carried out with the QUANTUM ESPRESSO software suite, which adopts ultrasoft pseudopotentials and density functional perturbation theory, and the cutoff energy is fixed at 60 Ry[45]. To improve the reliability of EPC results for the K-C system, appropriate *k*-point and *q*-point meshes are selected for diverse crystal configurations. The average contribution of each phonon mode to EPC effect is quantitatively evaluated, ensuring scientificity and rigor of computational data. The specific sampling grids are listed as follows: 10×10×2 *k*-points and 5×5×2 *q*-points for *C*2/*m* $K_7C$; 9×9×8 *k*-points and 3×3×4 *q*-points for *C*2/*m* $K_3C$; 10×10×8 *k*-points and 5×5×2 *q*-points for *C*2/*m* $K_2C$;

6×10×10 *k*-points and 3×5×5 *q*-points for *Imma* KC; 3×3×5 *k*-points and 3×3×5 *q*-points for *P*6/*mmm* $KC_2$; 16×16×6 *k*-points and 4×4×3 *q*-points for *C*2/*m* $KC_3$. Ultimately, the superconducting transition temperatures of K-C metallic compounds are calculated using the Allen-Dynes modified McMillan equation[46,47].

## 3. Results and discussion

### 3.1 Phase Diagram and Stability of K-C Compounds

To obtain stable compounds with exotic structural features in the K-C system, theoretical predictions were performed on a series of stoichiometric $K_xC_y$ compositions at 0 K and pressures of 1 atm, 10, 25, 50, 100, 200, and 300 GPa, where $x$ = 1, $y$ = 1-8; $x$ = 1-8, $y$ = 1; $x$ = 2, $y$ = 3,5; $x$ = 3, $y$ = 7.The structures with the lowest energy at each pressure were selected, and their average atomic formation energies relative to elemental potassium and solid carbon were calculated in eV/atom. A convex hull diagram of enthalpy was established based on all computational data. Thermodynamically stable K-C phases were marked with solid symbols, while unstable phases were denoted by hollow symbols. The results demonstrate that only $KC_8$ is thermodynamically stable at ambient pressure. KC and $KC_6$ become stable at 10 GPa. $K_7C$, $K_4C$, $K_2C$, KC and $KC_3$ gain thermodynamic stability at 25 GPa. Potassium-rich phases ($K_8C$, $K_3C$, $K_2C$) and carbon-rich phases (KC, $KC_2$) remain stable at 50 GPa. Only $K_3C$, KC and $KC_2$ maintain stability when the pressure rises to 100 GPa. It reveals that pressure acts as a critical factor governing the formation and thermodynamic stability of K-C compounds.

To further determine the stable pressure ranges of various K-C compounds, a pressure-composition phase diagram was constructed in this study. The results indicate that under ambient pressure, the experimentally reported *R*-3*m* type $KC_6$and *Fddd* type $KC_8$structures were successfully reproduced, verifying the reliability of the structural search method adopted in this work. As pressure increases gradually, a series of newly predicted K-C compounds including $K_8C$, $K_7C$, $K_4C$, $K_3C$, $K_2C$, KC, $KC_2$, and $KC_3$emerge on the convex hull of the phase diagram, and their stable pressure ranges are elaborated accordingly. Interestingly, pure phases are observed for $K_8C$, $K_7C$, $K_4C$, $K_3C$, $K_2C$, $KC_2$, and $KC_3$, while KC undergoes two successive phase transitions. The potassium-rich *C*2/*m*-$K_7C$ maintains thermodynamic stability from 20.0 GPa to 39.6 GPa. For the KC composition, the low-pressure stable phase is the $P6_3/mmc$ structure, which remains stable above 7.3 GPa. With further compression, the system experiences two structural phase transitions following the pathway $P6_3/mmc$→*Imma*→*R*-3*m*, with transition pressures of 21.7 GPa and 135.5 GPa respectively. As the carbon-richest phase, *C*2/*m* $KC_3$ is thermodynamically stable at 10.3–42.0 GPa.

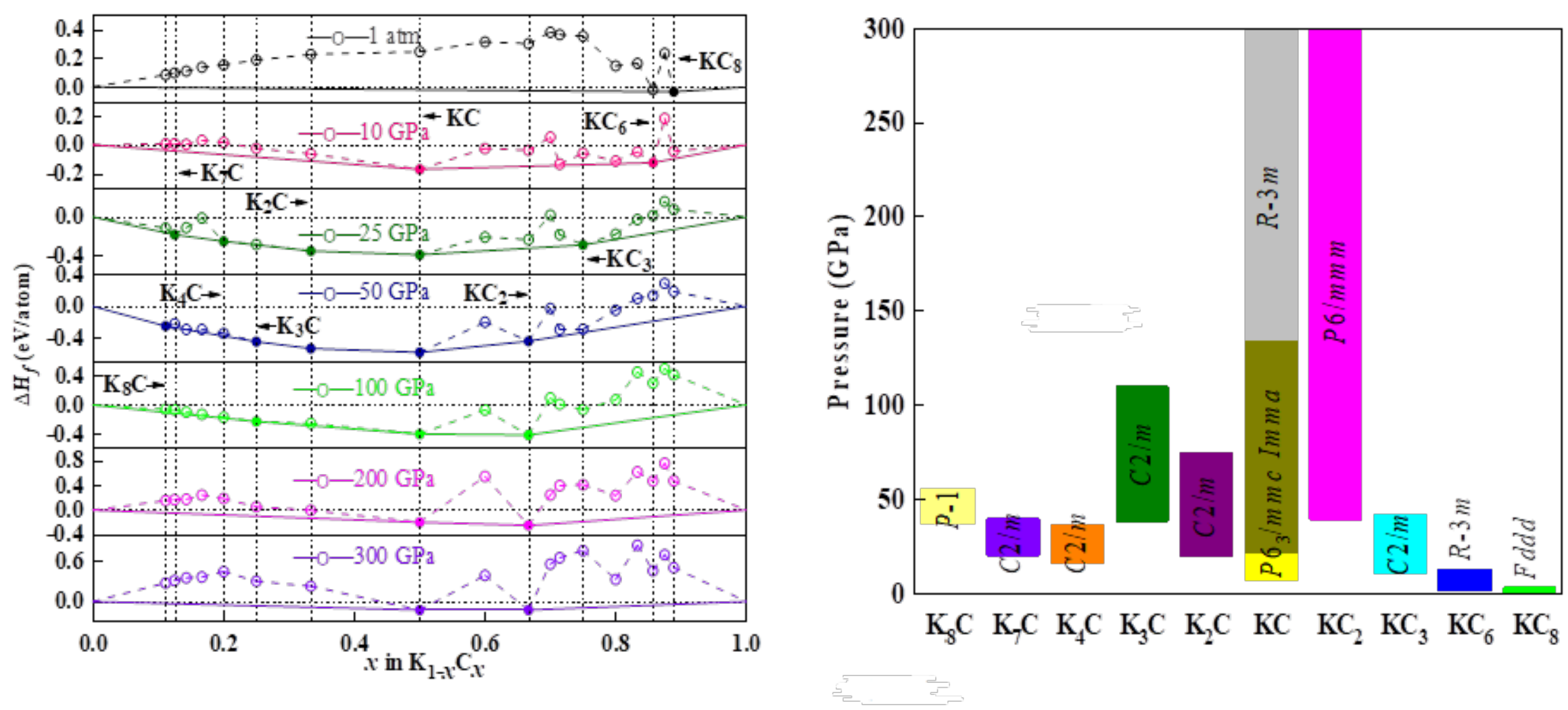

Figure 1 Stability of K-C Compounds. (a) Enthalpy of various K-C compounds relative to elemental K and C at 1 atm, 10, 25, 50, 100, 200, and 300 GPa. (b) Pressure-composition phase diagram of K-C compounds.

Besides verifying thermodynamic stability, the lattice dynamic stability was systematically evaluated by calculating phonon dispersion curves of all thermodynamically stable K-C compounds. As shown in the phonon dispersion spectra, no imaginary frequencies appear across the entire Brillouin zone, confirming that all thermodynamically stable K-C compounds possess favorable lattice dynamic stability. In addition, the phonon density of states of each stable structure was analyzed. The results reveal that phonon vibrations are predominantly contributed by potassium atoms in K-rich compositions, whose contribution far exceeds that of carbon atoms. By contrast, carbon atoms dominate the phonon vibration behavior in C-rich phases.

### 3.2 Crystal Structures of K-C Compounds

All ten newly discovered stable K-C compounds possess distinctive structural characteristics. As carbon content increases gradually, carbon atoms exist in diverse configurations including isolated single atoms, carbon dimers, zigzag chains, planar carbon structures, chain-like networks and folded carbon layers. Obvious differences are observed in space group types and atomic coordination numbers among various compounds.As the K-richest stable phase in the system, $K_8C$ adopts triclinic crystal system with $P$-1 space group. Its structure is constructed by interconnected $CK_{10}$ basic units. Each carbon atom is surrounded by ten potassium atoms. Carbon atoms remain isolated without direct bonding, carbon chain or layer aggregation, showing unique spatial features.$K_7C$ belongs to monoclinic system with $C2/m$ symmetry and consists of $CK_7$ units, where carbon atoms form dimers. Predicted $K_4C$ also has monoclinic $C2/m$ structure composed of $CK_7$ units with carbon dimer configuration.$C2/m$ $K_3C$ becomes stable under elevated pressure and contains carbon dimers constructed from $CK_9$ units. Structural variation occurs in monoclinic $C2/m$ $K_2C$, where carbon dimers evolve into zigzag chains based on $CK_5$ units.Stable KC at low pressure crystallizes in hexagonal *P*6₃/*mmc* symmetry. Carbon atoms form dimers, each coordinated with three potassium atoms, and the structure is assembled by $CK_3$ units.

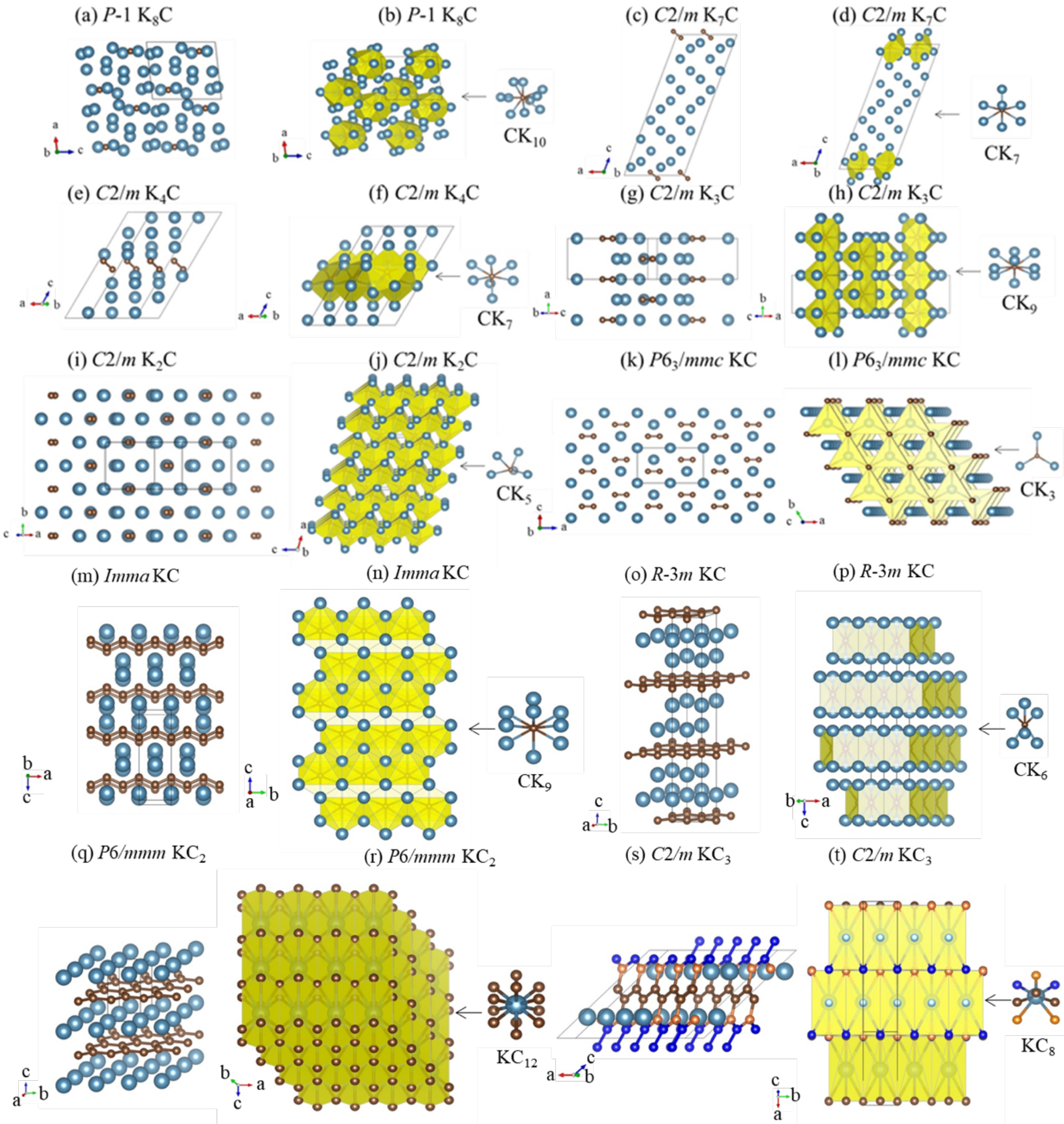

Figure 2 crystal structures of K-C compounds. (a) 50 GPa *P*-1 $K_8C$. (b) 50 GPa *P*-1 $K_8C$ and $CK_{10}$ unit. (c) 25 GPa *C*2/*m* $K_7C$. (d) 25 GPa *C*2/*m* $K_7C$ and $CK_7$ unit. (e) 25 GPa *C*2/*m* $K_4C$. (f) 25 GPa *C*2/*m* $K_4C$ and $CK_7$ unit. (g) 100 GPa *C*2/*m* $K_3C$. (h) 100 GPa *C*2/*m* $K_3C$ and $CK_9$ unit. (i) 50 GPa *C*2/*m* $K_2C$. (j) 50 GPa *C*2/*m* $K_2C$ and $CK_5$ unit. (k) 10 GPa $P6_3/mmc$ KC. (l) 10 GPa $P6_3/mmc$ KC and $CK_3$ unit. (m) 25 GPa *Imma* KC. (n) 25 GPa *Imma* KC and $CK_9$ unit. (o) 300 GPa *R*-3*m* KC. (p) 300 GPa *R*-3*m* KC and $CK_6$ unit. (q) 100 GPa *P*6/*mmm* $KC_3$. (r) 100 GPa *P*6/*mmm* $KC_2$ and $KC_{12}$ unit. (s) 25 GPa *C*2/*m* $KC_3$. (t) 25 GPa *C*2/*m* $KC_3$ and $KC_8$ unit.

Three stable phases of KC are identified, namely $P6_3/mmc$, *Imma* and *R*-3*m* phases in sequence with rising pressure. A phase transition occurs at 21.7 GPa, transforming KC into orthorhombic high-symmetry *Imma* phase. Each carbon atom is surrounded by nine potassium atoms, and the structure is assembled by $CK_9$ units. Meanwhile, carbon dimers evolve into zigzag carbon chains. Upon further compression up to 135.5 GPa, KC converts into trigonal *R*-3*m* phase with higher symmetry. Each carbon atom is coordinated with six potassium atoms and the structure consists of $CK_6$ units, where carbon atoms rearrange into planar configuration. This demonstrates that pressure exerts a decisive influence on carbon atomic arrangement. Carbon-rich *P*6/*mmm* $KC_2$ presents a typical $MgB_2$-type structure. Each potassium atom is surrounded by twelve carbon atoms, and carbon atoms form planar layers. In carbon-dominant *C*2/*m* $KC_3$, carbon atoms form folded layers, and the crystal structure is composed of $KC_8$ units.

**3.3 Chemical bonding of K-C compounds**

To further explore the chemical bonding characteristics and electron distribution rules of potassium-carbon compounds, this study systematically calculated and analyzed the electron localization function (ELF) of all stable potassium-carbon compound systems. Two-dimensional ELF can accurately distinguish different bonding types. When the ELF value exceeds 0.5, localized distribution of covalent bonds, lone electron pairs or core electrons exists in the system. An ELF value below 0.5 indicates ionic bonds. Three-dimensional ELF clearly depicts the spatial distribution of electron clouds around atoms, identifies covalent bonding regions and captures free electron distribution in lattice gaps of electrides, providing direct and reliable theoretical support for bonding identification. The two-dimensional and three-dimensional ELF of potassium-rich potassium-carbon compounds were calculated and analyzed. Obvious electron localization is observed in lattice gaps from 3D ELF patterns, proving these compounds possess fundamental electride properties. Combined analysis shows free electrons in gaps present zero-dimensional distribution, confirming all potassium-rich potassium-carbon compounds are zero-dimensional electrides.

In electron localization function (ELF) maps, the color gradient follows a distinct rule. Red areas represent high ELF values while blue areas stand for low values, with ELF values decreasing gradually from red to blue. Quantitative analysis of ELF maps reveals no obvious electron localization between potassium and carbon atoms, indicating ionic bonding between them. Remarkable electron aggregation is observed among carbon atoms, suggesting the formation of covalent bonds. In *C*2/*m* $K_7C$, the C-C covalent bonds almost disappear completely, leading to the collapse of the carbon framework. Electrons are mainly distributed among K atoms, presenting dominant metallic bonding characteristics, while C atoms exist as impurity or interstitial atoms. This system possesses the strongest metallicity and is a typical potassium-rich metallic phase. However, in *Imma* KC, The strength of C-C covalent bonds is lower than that in $KC_3$, and the polymerization degree of the carbon framework decreases. The interaction between K and C is strengthened. Apart from ionic bonds, weak covalent and polarization interactions also emerge. The overall bonding type transforms into a state dominated by ionic bonds with weak C-C covalent bonds, accompanied by enhanced metallicity. In contrast, Strong covalent bond networks are formed by carbon atoms in *C*2/*m* $KC_3$. The interaction between K and C is dominated by ionic bonds with nearly complete electron transfer, where K acts as cations and C constitutes the anionic framework. The system features a mixed bonding mode of covalent framework and ionic bonds, with C-C covalent interactions prevailing.

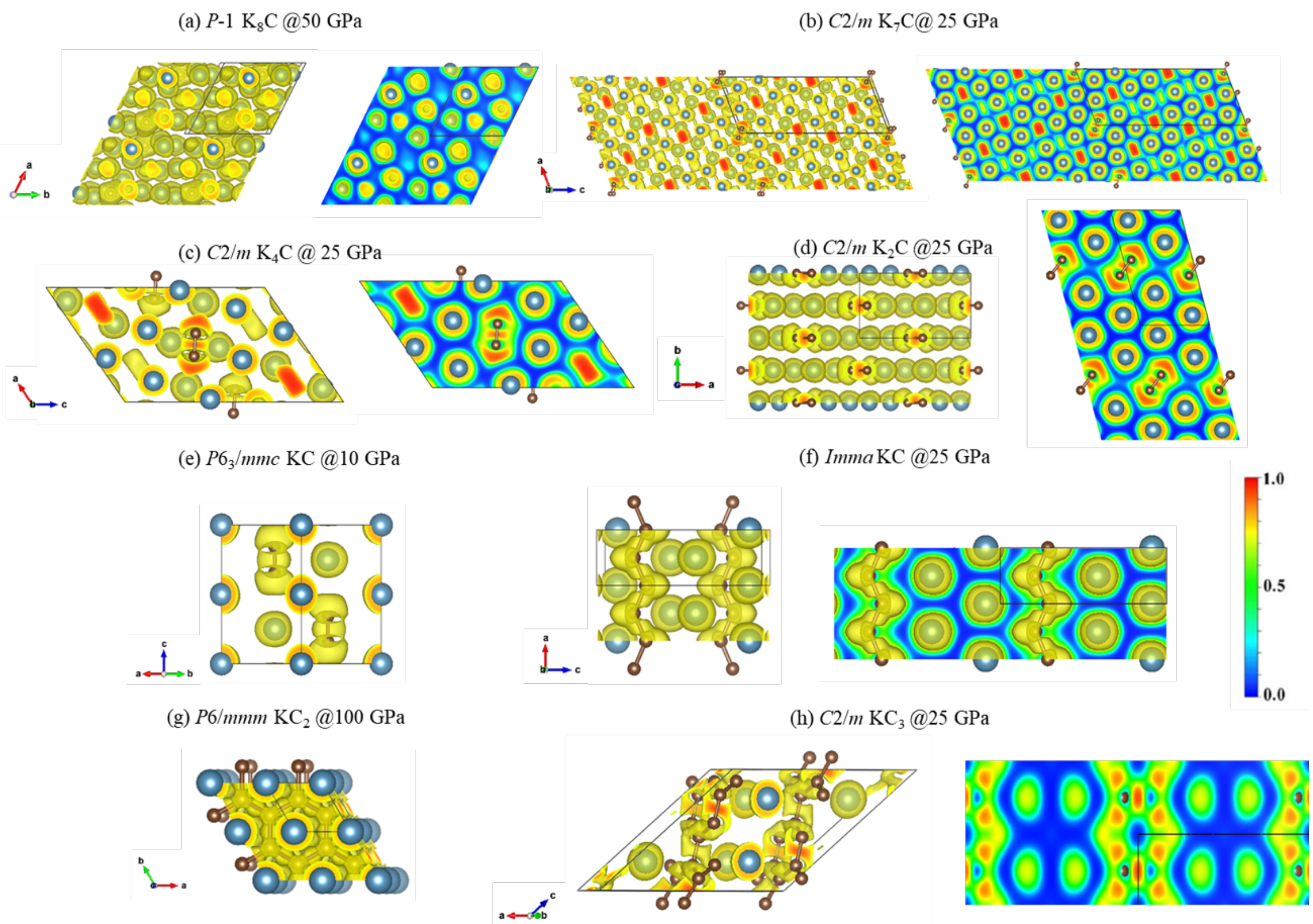

Figure 3 2D and 3D electron localization functions of potassium-carbon compounds with interstitial electrons. (a) ELF of *P*-1 $K_8C$ at 50 GPa. (b) ELF of *C*2/*m* $K_7C$ at 25 GPa. (c) ELF of *C*2/*m* $K_4C$ at 50 GPa. (d) ELF of *C*2/*m* $K_2C$ at 25 GPa. (e) ELF of *P*6$_3$/*mmc* KC at 10 GPa. (f) ELF of *Imma* KC at 25 GPa. (g) ELF of *P*6/*mmm* $KC_2$ at 100 GPa. (h) ELF of *C*2/*m* $KC_3$ at 25 GPa.

### 3.4 Electronic and Superconducting Properties of *C*2/*m* $K_7C$

No imaginary frequencies are observed in the phonon dispersion curves and projected phonon density of states of *C*2/*m* $K_7C$ at 25 GPa, confirming good dynamic stability of this phase. Phonon density of states analysis reveals that low-frequency vibrational modes (0–15 THz) are mainly contributed by K atoms, while C atoms exhibit localized vibrational modes only in the high-frequency range of approximately 45–50 THz. The distinct separation of vibrational frequencies indicates weak bonding between K and C atoms. Electronic band structure calculations show that bands cross the Fermi level without band gaps, demonstrating metallic characteristics. The relatively flat bands near the Fermi level correspond to large electron effective mass, which facilitates enhanced electron-phonon coupling. The projected density of states indicates that the density of states at the Fermi level is predominantly derived from K 3*d* orbitals, while C 2*p* orbitals mainly contribute to the valence bands. This suggests that the electrical conduction is dominated by K atoms, and C atoms exist in a weakly bonded localized state.

Overall, the metallicity, dynamic stability, non-zero density of states at the Fermi level and K-dominated low-frequency vibrations of *C*2/*m* $K_7C$ are favorable for conventional phonon-mediated superconductivity. Nevertheless, due to the weak covalent interaction within the carbon framework, the superconducting transition temperature of this phase is expected to be lower than that of carbon-rich phases in the K-C system.

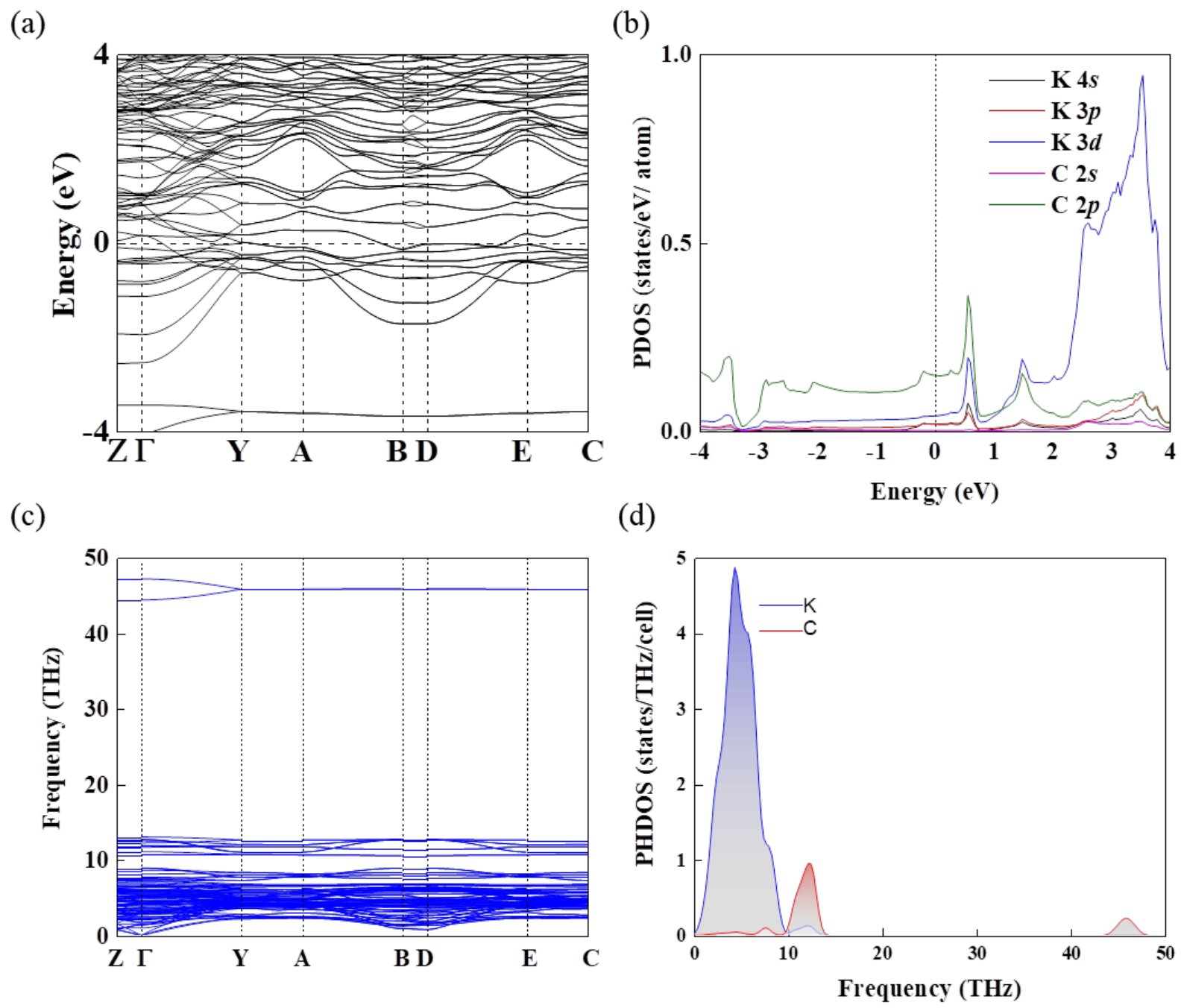

Figure 4 Superconducting properties of *C*2/*m* $K_7C$ at 25 GPa. (a) Electronic Band Structure. (b) projected density of states (PDOS). (c) Phonon dispersion curve. (d) Phonon projected density of states (PHDOS).

### 3.5 Superconducting properties of *Imma* KC and *C*2/*m* $KC_3$

Combined with previous analysis, the *Imma*-phase KC compound exhibits strong electron-phonon coupling characteristics. Based on BCS theory, the Coulomb pseudopotential $\mu^*$ is set to 0.1. The calculated superconducting critical temperature of Imma-phase KC reaches 21.4 K at 25 GPa, remarkably higher than reported values of graphite intercalation compound $KC_6$ (0.1 K) and potassium fulleride $K_3C_{60}$ (18 K). Analysis of electron-phonon coupling integral parameter $\lambda(\omega)$ and phonon projected density of states reveals that the electron-phonon coupling constant $\lambda$ stems from two vibration contributions. Low-frequency phonon vibrations below 14 THz originate from potassium atomic motion, while high-frequency vibrations above 14 THz are dominated by carbon atoms. Electrons from K 3*p* and C 2*p* orbitals play a dominant role in the coupling process. The electron-phonon coupling constant $\lambda(\omega)$ of *Imma*-phase KC is 0.74 at 25 GPa, exceeding that of $MgB_2$ ($\lambda$ = 0.61), which accounts for its outstanding superconducting performance. Pressure exerts obvious effects on superconducting transition temperature. With increasing pressure, the logarithmic average phonon frequency rises while the electron-phonon coupling constant declines. The competitive effect of the two factors leads to a marked drop in superconducting transition temperature.

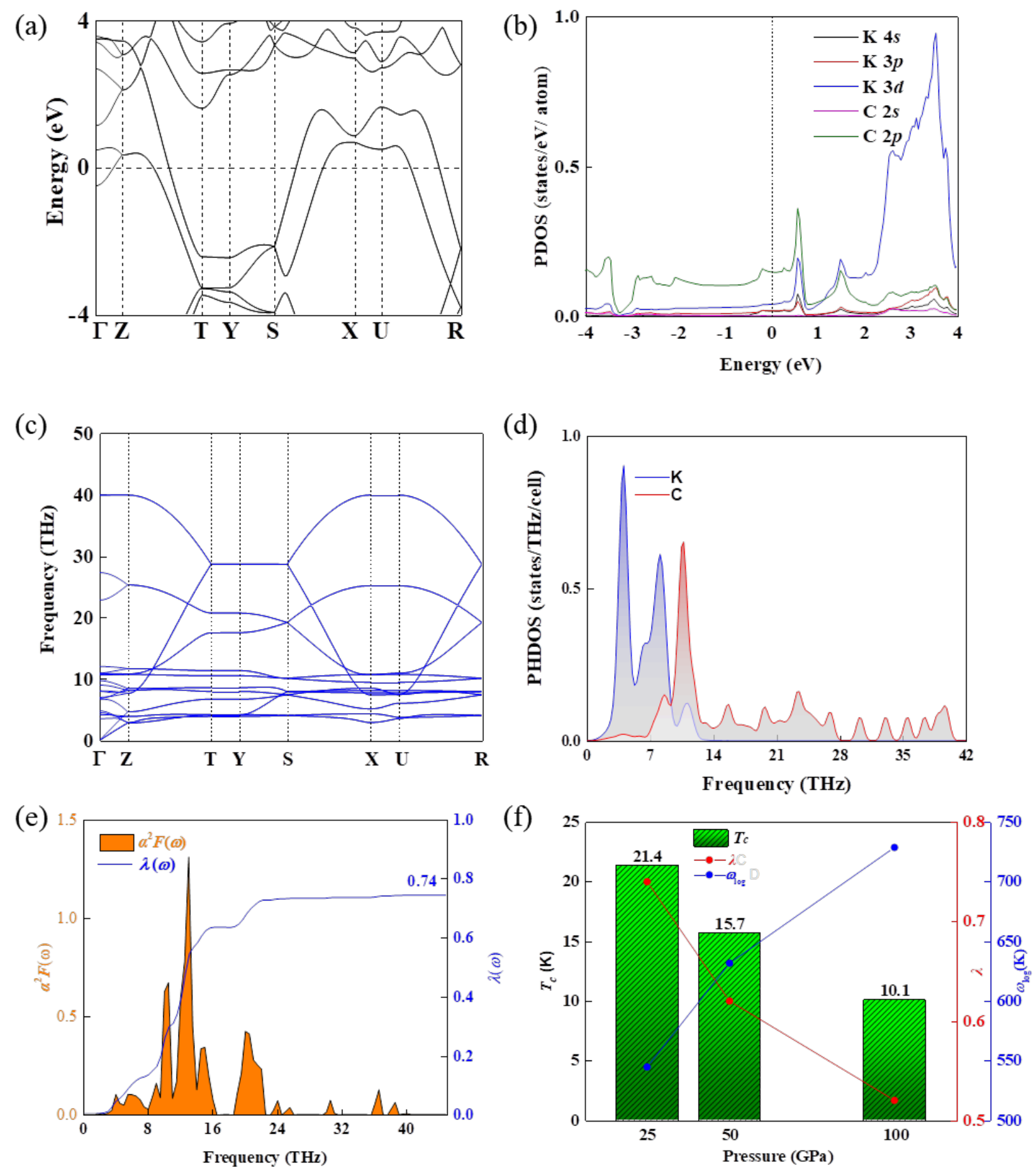

Figure 5 Superconducting properties of *Imma* KC at 25 GPa. (a) Electronic Band Structure. (b) projected density of states (PDOS). (c) Phonon dispersion curve. (d) Phonon projected density of states (PHDOS). (e) Dependence of Eliashberg spectral function $\alpha^2F(\omega)$ and electron-phonon coupling constant $\lambda(\omega)$ on frequency. (f) Variations of superconducting transition temperature $T_c$, electron-phonon coupling constant $\lambda$ and logarithmic average phonon frequency $\omega_{\log}$ with pressure.

Compared with *Imma*-phase KC, *C*2/*m*-phase $KC_3$ has a slightly lower superconducting transition temperature at 25 GPa, due to intrinsically different superconducting mechanisms.Electron-phonon coupling analysis reveals that its coupling strength is mainly contributed by medium and high-frequency phonons within 10–50 THz. Low-frequency vibrations below 10 THz dominated by K atoms contribute barely, which is distinctly different from *Imma*-phase KC.The superconducting transition temperature of *C*2/*m*-phase $KC_3$ decreases slightly when pressure rises to 50 GPa. Increasing pressure reduces the electron-phonon coupling constant $\lambda$ and raises the logarithmic average phonon frequency $\omega_{\log}$, forming a competitive effect. The decline of coupling strength dominates the change of critical temperature, which accords with the trend of hydrogen-rich superconductors like $YH_{10}$ and $Li_2MgH_{16}$. In addition, the superconducting transition temperatures of other potassium-carbon compounds were calculated. The detailed values are listed in Table 1, providing fundamental data for subsequent relevant studies.

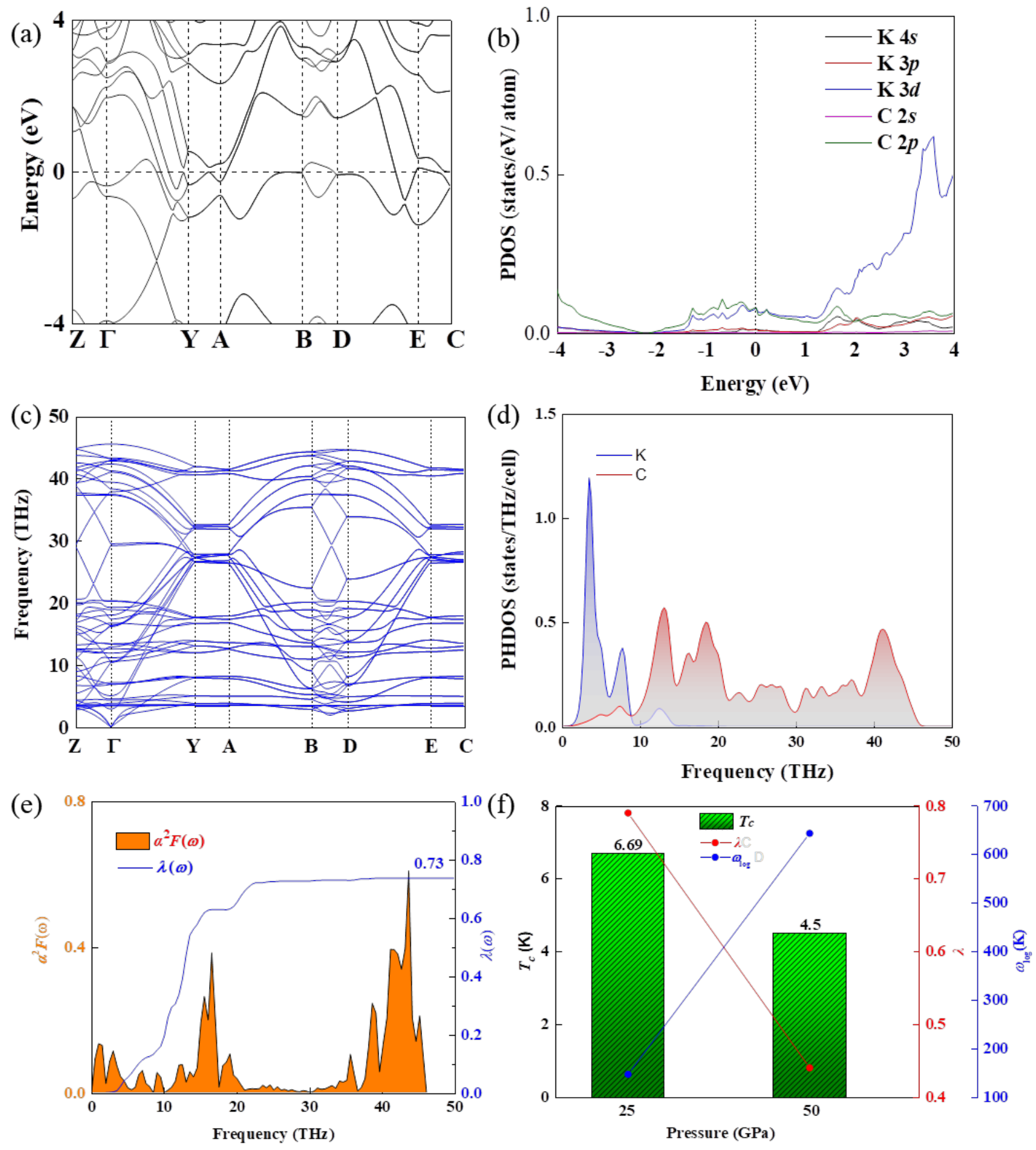

Figure 6 Superconducting properties of $C2/m$ $KC_3$ at 25 GPa. (a) Electronic Band Structure. (b) projected density of states (PDOS). (c) Phonon dispersion curve. (d) Phonon projected density of states (PHDOS). (e) Dependence of Eliashberg spectral function $\alpha^2F(\omega)$ and electron-phonon coupling constant $\lambda(\omega)$ on frequency. (f) Variations of superconducting transition temperature $T_c$, electron-phonon coupling constant $\lambda$ and logarithmic average phonon frequency $\omega_{\log}$ with pressure

Table 1 Superconducting properties of predicted K-C compounds.

| Phase | Pressure (GPa) | $\lambda$ | $\omega_{\log}$ (K) | $N(E_f)$ (states/Ry/cell) | $T_c$ （K） |
|---|---|---|---|---|---|
| *C*2/*m* $K_7C$ | 25 | 0.40 | 134.6 | 52.68 | 0.6 |
| *C*2/*m* $K_3C$ | 50 | 0.29 | 320.9 | 18.45 | 0.1 |
| *C*2/*m* $K_3C$ | 100 | 0.26 | 327.4 | 11.25 | 0.04 |
| *C*2/*m* $K_2C$ | 25 | 0.33 | 292.8 | 13.85 | 0.3 |
| *C*2/*m* $K_2C$ | 50 | 0.26 | 353.6 | 12.26 | 0.04 |
| *Imma* KC | 25 | 0.74 | 544.7 | 5.97 | 21.4 |
| *Imma* KC | 50 | 0.62 | 631.7 | 5.23 | 15.7 |
| *Imma* KC | 100 | 0.52 | 728.7 | 4.53 | 10.1 |
| *P*6/*mmm* $KC_2$ | 50 | 0.25 | 787.1 | 2.78 | 0.04 |
| *P*6/*mmm* $KC_2$ | 100 | 0.24 | 877.3 | 2.73 | 0.02 |
| *C*2/*m* $KC_3$ | 25 | 0.79 | 146.6 | 8.90 | 6.69 |
| *C*2/*m* $KC_3$ | 50 | 0.44 | 643.2 | 8.39 | 4.5 |

## 4. Conclusion

This paper combines swarm intelligence search strategy with density functional theory to systematically explore structures and properties of binary K-C system under high pressure. A series of new stable potassium-carbon phases with diverse stoichiometries including $K_8C$, $K_7C$, $K_4C$, $K_3C$, $K_2C$, KC, $KC_2$ and $KC_3$ are identified. The pressure-composition phase diagram of K-C system is established to illustrate stability ranges of each phase under varying pressures.$K_8C$, $K_7C$ and $K_4C$ are verified as electrides with interstitial quasi-atoms donated by potassium atoms in lattice gaps, which determine their electronic behaviors. At 25 GPa, *C*2/*m*-phase $K_7C$ is an electride superconductor with critical temperature of 0.6 K. *Imma*-phase KC possesses superior superconductivity with transition temperature up to 21.4 K, serving as a promising high-performance superconducting material. Carbon-rich *C*2/*m*-phase $KC_3$ also exhibits superconductivity with $T_c$ of 6.69 K under the same pressure. Electron-phonon coupling constant and logarithmic average phonon frequency vary oppositely with increasing pressure. Such competitive effect causes gradual decline of superconducting transition temperature of $KC_3$.The findings expand the database of binary metal carbides and deepen understanding of structure-property correlation of potassium-carbon compounds. It provides valuable theoretical basis and practical guidance for structural design, performance modulation and superconducting mechanism research of binary metal carbides.

**Acknowledgments**

The authors acknowledge the funding support from the National Natural Science Foundation of China (Grant No. 11704062) and the "111 Center" (Grant No. B25030).